\documentclass[twocolumn,prl,aps,showpacs,superscriptaddress]{revtex4}

\usepackage{graphicx}

\def\be{\begin{equation}}
\def\ee{\end{equation}}
\def\ba{\begin{eqnarray}}
\def\ea{\end{eqnarray}}

\def\LSCO{La$_{2-x}$Sr$_x$CuO$_4$ }

\def\124{YBa$_2$Cu$_4$O$_8$ }

\def\C60{A$_x$C$_{60}$ }

\begin{document}

\title{A proximity induced pseudogap - evidence for preformed pairs}

\author{Ofer Yuli}
\affiliation{Racah Institute of Physics, The Hebrew University of
Jerusalem, Jerusalem 91904, Israel}

\author{Itay Asulin}
\affiliation{Racah Institute of Physics, The Hebrew University of
Jerusalem, Jerusalem 91904, Israel}

\author{Yoav Kalcheim}
\affiliation{Racah Institute of Physics, The Hebrew University of
Jerusalem, Jerusalem 91904, Israel}

\author{Gad Koren}
\affiliation{Department of Physics, Technion - Israel Institute of
Technology, Haifa 32000, Israel}


\author{Oded Millo} \email{milode@vms.huji.ac.il}
\affiliation{Racah Institute of Physics, The Hebrew University of
Jerusalem, Jerusalem 91904, Israel}

\begin{abstract}

The temperature evolution of the proximity effect in Au/\LSCO and
La$_{1.55}$Sr$_{0.45}$CuO$_4$/\LSCO bilayers was investigated
using scanning tunneling microscopy. Proximity induced gaps,
centered at the chemical potential, were found to persist above
the superconducting transition temperature, $T_c$, and up to
nearly the pseudogap crossover temperature in both systems. Such
independence of the spectra on the details of the normal metal cap
layer is incompatible with a density-wave order origin. However,
our results can be accounted for by a penetration of incoherent
Cooper pairs into the normal metal above $T_c$.

\end{abstract}

\pacs{74.78.Fk, 74.72.Dn, 74.50.+r, 74.78.Bz, 74.25.Jb}

\maketitle

Since the discovery of the pseudogap (PG) phase in cuprate high
temperature superconductors, it has been deemed that its origin
holds key information regarding the pairing mechanism in these
materials \cite{Cho, Pines}. Despite the intensive study of the
subject \cite{Timusk, PG Rev}, it remains unclear whether the PG
phenomenon is related to superconductivity or not. The debate on
the origin of the PG revolves around the following two paradigms:
1) It is a superconducting precursor state where electrons pair
incoherently above $T_c$. 2) The PG bears no direct relation to
superconductivity and is a manifestation of a competing order
setting in at $T^* > T_c$. The former is based on many reports
claiming that the transition temperature for pairing exceeds the
phase locking temperature, at least in the underdoped regime
\cite{Uemura,Kivelson,Corson,Ong-Nernst}. The second paradigm
points to the various ordered states predicted to compete with
superconductivity in strongly correlated systems \cite{Chakra,
Varma, Li} like the cuprates. One way to distinguish between the
two is to examine whether unique spectral properties associated
with pairing are present in the PG regime. Concomitantly, a recent
study of the photo-emission spectra in Bi2212 single crystals
\cite{Kanigel}, reported that a Bogoliubov-like dispersion curve,
typical of superconductivity, was measured above $T_c$ in the
gapped anti-nodal region. The minimal energy value of the leading
edge in the energy distribution curve, consistently found at
k$_F$, led the authors to attribute their findings to a finite
pairing amplitude existing above $T_c$ rather than to a density
wave ordering (for which the gap minimum does not necessarily
reside at $E_F$).

Another distinctive spectral fingerprint of the superconducting
order parameter is its ability to penetrate an adjacent normal
metal via Andreev reflections, a process known as the proximity
effect: When a normal metal (N) and a superconductor (S) are
placed in good electrical contact with one another, a hole-like
quasiparticle with energy smaller than the superconductor gap,
$\Delta$, impinging on the N/S interface from the N side, may be
retro-reflected as an electron, whilst destroying a Cooper pair on
the S side. This unique process (that can be viewed as a transfer
of a Cooper pair from S to N) induces superconducting correlations
in N, which are manifested as a gap in the density of states (DOS)
of the normal metal \cite{ASJ review}. Experimentally, the nature
of the proximity effect has been studied mainly at temperatures
below $T_c$. An investigation of the proximity effect above $T_c$
in the PG temperature regime is still lacking and is the focal
point of the present study.

In this letter we report the temperature evolution of the
proximity induced gap in correlation with sample morphology
measured on Au/\LSCO and La$_{1.55}$Sr$_{0.45}$CuO$_4$/\LSCO
bilayers using scanning tunneling microscopy (STM) and
spectroscopy (STS). Our main finding, observed on both
configurations, is a smooth evolution of the superconductor
proximity gap into a proximity induced PG, as the temperature was
raised above $T_c$ for bilayers comprising an underdoped \LSCO
[LSCO(x)] film. The induced gap gradually filled up by spectral
weight until it became undetectable very close to the doping
dependent T$^*$ value we found via STS for the corresponding bare
LSCO film \cite{Yuli-PRB}. The magnitude of the proximity gaps
exhibited a similar spatial decay as a function of distance from
an \emph{a}-axis facet below and above $T_c$, suggesting a common
origin. In contrast, in bilayers comprising an overdoped LSCO
layer, the proximity gap disappeared close to $T_c$. We claim that
the observed independence of the $T > T_c$ proximity-gap features
on the properties of the normal-metal layer (\emph{e.g.}, its
Fermi surface), in particular the fact that the induced gap is
always centered at the Fermi energy, $E_F$, is inconsistent with a
density wave origin for the induced gap. On the other hand, these
same results, as well as the spatial dependence of the gap size
found above $T_c$, are indicative of a finite pairing amplitude
present at $T > T_c$.

N/LSCO(x) bilayers (N = Au and La$_{1.55}$Sr$_{0.45}$CuO$_4$) with
$x$ = 0.08, 0.10 (underdoped) and $x$ = 0.18 (overdoped) were
epitaxially grown on (100)SrTiO$_3$ wafers by laser ablation
deposition with \emph{c}-axis orientation perpendicular to the
substrate [see schematic illustration in Fig. 1(a)]. The LSCO(x)
films were 90 nm thick, and the N cover-layer, grown \emph{in
situ} without breaking the vacuum, was $\sim$ 7 nm thick. The
\emph{in situ} metallic coating prevented the rapid surface
degradation known to occur in LSCO films and enabled high quality
STM topographic imaging. The underlying LSCO surface morphology
comprised square grains of $\sim$ 100 nm lateral size and 10-20 nm
in height, exposing relatively large \emph{a}-axis facets (see
Fig. 1). The gold layer exhibited a granular morphology with a
typical grain size of $\sim$ 5 nm and roughness of 1-2 nm [see
blow up in Fig. 1(c)]. The bulk superconducting transition
temperatures were obtained by 4-probe resistance vs. temperature
[$R(T)$] measurements and are presented in Fig. 1(b). We have also
measured the properties of a bare 90~nm
La$_{1.55}$Sr$_{0.45}$CuO$_4$ film. The $R(T)$ data showed no sign
of a superconducting transition down to 4~K and the tunneling
spectra taken at 4.2~K, exhibited gapless Ohmic behavior.
Therefore, we conclude that the $x = 0.45$ layer was metallic in
the temperature range of our experiments ($T
> 4.2$ K).

\begin{figure}
\includegraphics[width=2.6in]{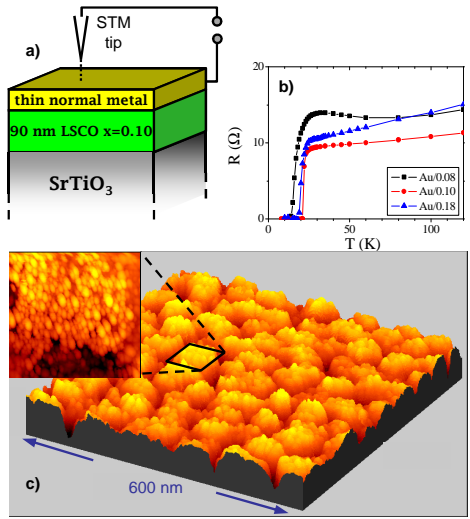}
\caption{(Color online) a) Schematic of the experimental setup. b)
R(T) of the Au/\LSCO bilayers with x = 0.08, 0.10 and 0.18. c) 600
x 600 nm$^2$ STM topographic image of an
Au/La$_{1.90}$Sr$_{0.10}$CuO$_{4}$ bilayer. The small granular
gold is clearly seen on top of the underlying LSCO crystallites.}
\label{fig1}
\end{figure}

\begin{figure}[b]
\includegraphics[width=2.2in]{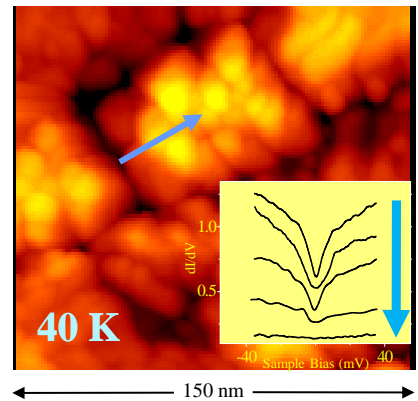}
\caption{(Color online) 150 x 150 nm$^2$ topographic image of an
Au/LSCO($x = 0.10$) taken at 40 K. The Au coverage is seen as the
grainy surfce on top of the square LSCO crystallite. Inset: dI/dV
vs. V curves taken along the arrow indicated in the topographic
image. The gap size was found to depend on the distance from the
grain edges in a similar manner both below and above $T_c$. The
curves are vertically shifted for clarity.} \label{fig2}
\end{figure}

Our tunneling spectra, namely dI/dV vs. V curves, acquired by
momentarily disabling the feedback loop, were obtained at specific
well defined locations, thus revealing the local DOS in
correlation with the surface topography. In Fig. 2 we present a
topographic image taken on an Au/LSCO($x=0.10$) bilayer at 40 K,
well within the PG temperature regime ($T^{onset}_c$ = 26 K). The
inset shows a set of dI/dV vs. V curves measured consecutively
along the blue arrow, depicting the dependence of the measured gap
size on the distance from the LSCO grain boundary, namely from the
interface between an \emph{a}-axis LSCO facet and the Au layer. As
the tip was moved away from the LSCO grain edge, the gap size
diminished until near the grain center, far enough from the grain
edge, the spectra turned gapless featuring a metallic
structureless DOS. Importantly, a similar spatial dependence was
found below $T_c$ at $\sim$ 4.2 K. The spatial dependence of our
spectra can be understood within the framework of an anisotropic
proximity effect expected for \emph{d}-wave superconductors, as
reported by Sharoni \emph{et al.} \cite{Sharoni caxis} for
Au/YBa$_2$Cu$_3$O$_7$ bilayers: the proximity effect takes place
predominantly at N/S interfaces involving \emph{a}-axis oriented
facets, \emph{i.e.} at the grain boundaries. The spatial evolution
of the induced gap showed a behavior similar to the conventional
proximity effect, or, in other words, as the STM tip was moved
away from the side facet, toward the grain center, a monotonic
decay of the gap size was measured. The spectra then turned Ohmic
at distances larger than the penetration length which in the dirty
limit is given by the expression $\xi_N = (\frac{\hbar
v_{FN}l_N}{6 \pi K_BT}) ^{1/2}$. In our case, $\xi_N \approx$ 10 -
20 nm at 4.2 K (taking $v_{FN} = 1.4 \cdot10^{6}$ m/s and the mean
free path $l_N$ to be the average Au grain-size, $\sim$ 5 nm)
which accounts for the Ohmic behavior on top of the grain about 50
nm away from the grain edge \cite{Sharoni caxis}. Such behavior
agrees qualitatively with the predicted evolution of the DOS
calculated by L\"{o}fwander \cite{Lofwander}. As the temperature
was raised, areas exhibiting Ohmic spectra inhabited a larger
fraction of the surface. Nevertheless, areas exhibiting pronounced
induced gaps were still found as in the 4.2 K case, predominantly
near grain edges.

\begin{figure}[b]
\includegraphics[width=2.7in]{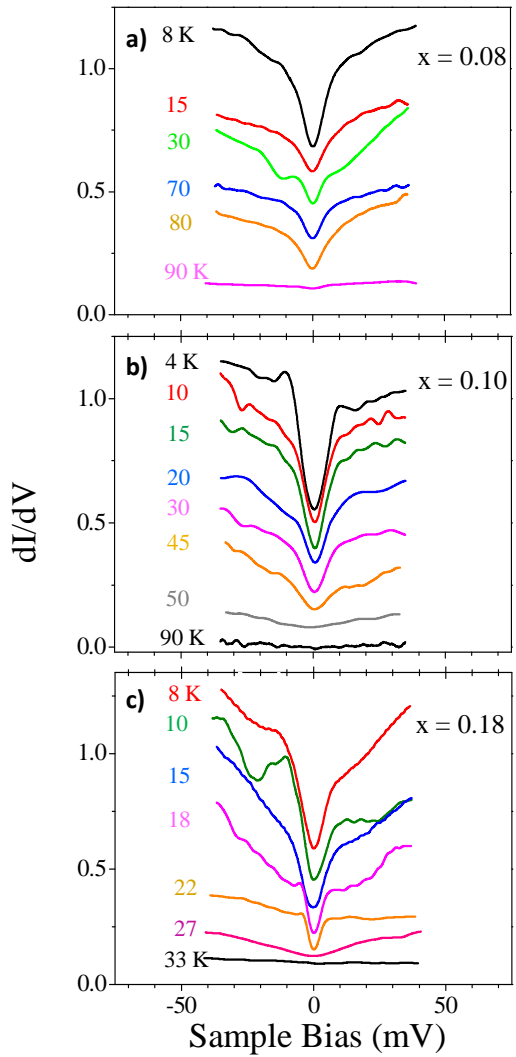}
\caption{Temperature evolution of the local DOS measured on the
Au/LSCO(x) bilayer for a) $x = 0.08$, b) $x = 0.10$ and c) $x =
0.18$. The conventional proximity gap found below $T_c$ evolved
smoothly to an Ohmic spectra at T $\approx$ T$^*$ of the
corresponding bare LSCO for both $x = 0.08$ and $x = 0.10$ and
disappeared at $T \approx T_c^{onset}$ for the $x = 0.18$ bilayer.
The curves are vertically shifted for clarity, and normalized to
the gap edge value. The vertical axis corresponds to the curve
taken at 8 K in a) and c) and at 4.2 K in (b).} \label{fig3}
\end{figure}

A detailed description of the induced-gap's temperature dependence
is portrayed in Fig. 3, which presents tunneling spectra acquired
near grain edges in Au/LSCO(x) bilayers with $x = 0.08, 0.10$ and
$0.18$. To ensure the gaps were proximity gaps, a correlation with
local topography was done (similar to the case presented in Fig.
2) thus avoiding confusion with data originating from possibly
local low Au coverage. The transition onset temperatures of the
bilayers were 24 K, 26 K and 27 K for the $x = 0.08$, $0.10$ and
$0.18$, respectively, as seen in Fig. 1(c). Nevertheless, and this
is the central result of the presented report, the proximity gap
structure did not vanish at $T_c$ for the underdoped bilayers.
Moreover, no noticeable change in the measured spectra was
observed near T $\sim$ $T_c$ for both underdoped samples. Rather,
the gap seems to smoothly fill up, until it disappears completely
at T($x=0.08$) $\sim$ 90 K and T($x = 0.10$) $\sim$ 50 K in
striking resemblance to T$^*$ values we measured by STS for the
corresponding bare underdoped LSCO films \cite{Yuli-PRB}. In
contrast, no proximity induced gap was found above $T_c$ in the
spectra of the $x = 0.18$ overdoped bilayers [see Fig. 3(c)], in
accordance with reports of a conventional normal state in the LSCO
overdoped regime \cite{Yuli-PRB, Ino}. We have also measured an
underdoped (x=0.08) and overdoped ($x = 0.18$) LSCO, coated by 7
nm of La$_{1.55}$Sr$_{0.45}$CuO$_4$ which replaced the Au as the
normal metal. The results were essentially very similar - the gap
remained centered at zero bias and disappeared at $\sim$ 90 K for
the underdoped bilayer and at $T_c$ for the overdoped layer.

The similar spatial dependence (as a function of distance from the
\emph{a}-axis facets) exhibited by both the proximity induced PG
($T > T_c$) and the superconductor proximity gap ($T < T_c$),
suggests that the PG phase is closely related to
superconductivity, \emph{e.g.} that it may be associated with the
existence of uncondensed Cooper pairs above $T_c$. Nevertheless,
the possible induction of a proximity gap by competing orders
suggested for the PG cannot be ruled out \emph{a-priori}. These
phases, which do not involve Cooper pairing, are generally
characterized by an order parameter with non-zero momentum
center-of-mass, \textbf{Q}. In light of our results we need to
address the question: can an order parameter with \textbf{Q}
$\neq$ 0 induce a gap in a normal metal and if so will its
properties comply with our data? We shall first focus for clarity
on the phases induced by a nesting condition such as the charge
and spin density wave. In a N/DW (DW - density-wave of some sort)
configuration, normal-metal quasiparticles with sub-gap energy
impinging on the gapped DW interface, change their momenta by the
wave vector \textbf{Q} of the DW pattern in an unconventional
reflection process termed a \textbf{Q} reflection \cite{Q,Bauer}.
Consequently, a zero charged particle-hole entity with the
transferred momentum \textbf{Q} is created in the DW side. The
nesting condition E(\textbf{k}$\pm$\textbf{Q}) = E(\textbf{-k})
responsible for the gap in the DW condensate, induces a sign
change of the reflected electron's velocity, at least in the
semiclassical limit where \textbf{v} =
$\partial$E(\textbf{k})/$\partial$\textbf{k}. The subsequent mix
of \textbf{k} states with \textbf{k}$\pm$\textbf{Q} in the normal
metal can be viewed as a penetration of DW correlations into N.
Such a mix inevitably yields a gap centered at the corresponding
energy E(\textbf{k}), as was discussed in detail by Kanigel
\emph{et al}. \cite{Kanigel} for the cuprate normal state. For the
special case of \textbf{Q} = 2\textbf{k}$_F$, the induced gap will
open at $E_F$ in accordance with our results. However, in view of
this proposed scenario we need to recall that the gap remained
centered at zero bias when the Au was replaced by a different
normal metal, La$_{1.55}$Sr$_{0.45}$CuO$_4$, albeit now obviously
\textbf{Q} $\neq$ 2\textbf{k}$_F$. Such robust behavior of the gap
location implies the induced gap in N does not stem from a
non-zero momentum order. Note that the latter argument does not
require the nesting condition assumed above. Any phase which mixes
\textbf{k} and \textbf{k}$\pm$\textbf{Q} states in N (\emph{e.g.}
a \emph{d}-density-wave order) will eventually induce a gap
removed from the chemical potential upon changing the Fermi
surface in N. In contrast, Andreev reflections, unlike \textbf{Q}
reflections, involve particle-hole mixing with zero momentum
(\textbf{Q} = 0) and accordingly, the induced gap will
\emph{always} be centered at $E_F$ regardless of the normal metal
details. We thus conclude that our findings comply better with the
pre-formed pairs scenario for the PG compared to any of the
suggested competing orders.

The question now is, whether a system comprising Cooper pairs
lacking global phase-coherence can undergo an Andreev reflection.
The answer entails a solution of the Bogoliubov-de Gennes
equations in the presence of thermal phase fluctuations as was
done by Franz and Millis \cite{F-M}. Following the framework put
forward in Ref. \onlinecite{Kivelson}, the normal state was
modelled as a plasma of unbound vortex-antivortex pairs
(associated with the Kosterlitz-Thouless transition). Each vortex
is surrounded by a circulating supercurrent which leads to a
Doppler shift in the local quasiparticle excitation spectrum of
$\Delta$E = $\hbar$\textbf{k}$\cdot$\textbf{v}$_s$ where
\textbf{v}$_s$ is the local superfluid velocity, which is related
to the order parameter phase by
\textbf{v}$_s$(\textbf{x})=$\hbar\nabla\theta$(\textbf{x})$/$2m.
The local DOS was calculated by averaging over the phase
fluctuations assumed to vary in space slower than the
superconductor coherence length, $\xi_s$. Choi \emph{et al.} \cite
{Choi} extended this calculation to the PG regime and predicted an
enhancement of the differential conductivity due to Andreev
reflected Cooper pairs which lack long-range phase coherence. This
work established the finite probability for the occurrence of
Andreev reflections by preformed pairs. We note, however, that the
effect of such reflections on the DOS in the normal metal has not
been calculated yet.

It should be noted that in a previous Andreev spectroscopy study
of LSCO in the PG regime, no surface Andreev bound states were
found above $T_c$ \cite{Dagan}. This observation may appear to
contradict ours. However the conditions for the formation of these
bound states, a process that involves alternating Andreev and
normal reflections \cite{Hu,TK}, are probably more stringent
compared to gap induction which requires a single Andreev
reflection. This notion is supported by the commonly observed
disappearance of the zero-bias conductance peak \cite{ASJ review}
associated with the Andreev bound states even at $T < T_c$
\cite{Dagan,Greene2}.

In conclusion, we have observed a proximity gap above $T_c$ in the
DOS of a normal metal over-coating an underdoped LSCO film. The
induced gap survives up to a temperature close to the PG onset
temperature $T^*$, suggesting a common origin for both phenomena.
The similar spatial dependence of the proximity induced PG and the
superconductor proximity gap indicate that the origin of the PG is
related to superconductivity. Moreover, the gap minimum was always
measured to be at $E_F$, even for two very different N layers, Au
and La$_{1.55}$Sr$_{0.45}$CuO$_4$, a finding difficult to
reconcile with a PG order parameter characterized by \textbf{Q}
$\neq$ 0. The predicted ability of Cooper pairs to induce
correlations in a normal metal in the presence of (coherence
breaking) phase-fluctuations leads us to ascribe the PG to a state
consisting of Cooper pairs with short-range phase coherence.

The authors are grateful to G. Vachtel, A. Frydman, A. Kanigel, G.
Deutscher and S. Kivelson for invaluable discussions. We
especially thank D. Orgad for his contributions and stimulating
talks. This research was supported in parts by the German Israeli
DIP project, the Israel Science Foundation, the Center of
Excellence Program (grant No. 1565/04) and the United States -
Israel Binational Science Foundation (grant No. 2004162). O. M.
thanks support from the Harry de Jur Chair in Applied Science and
G. K. from the the Heinrich Hertz Minerva Center for HTSC, the
Karl Stoll Chair in advanced materials and by the Fund for the
Promotion of Research at the Technion.

\end{document}